\documentclass[aip,jcp,preprint,preprintnumbers,amsmath,amssymb,floatfix,showpacs]{revtex4-1}
\usepackage{graphicx}
\usepackage{amsfonts,amsbsy}
\begin{document}

\title{Quantum Monte Carlo study of the first-row atoms and ions}
\author{P.\ Seth}
\email{ps479@cam.ac.uk}
\affiliation{Theory of Condensed Matter Group, Cavendish Laboratory,
University of Cambridge, J.~J.~Thomson Avenue, Cambridge CB3 0HE,
United Kingdom}
\author{P.\ L\'opez R\'{\i}os}
\affiliation{Theory of Condensed Matter Group, Cavendish Laboratory,
University of Cambridge, J.~J.~Thomson Avenue, Cambridge CB3 0HE,
United Kingdom}
\author{R.~J.\ Needs}
\affiliation{Theory of Condensed Matter Group, Cavendish Laboratory,
University of Cambridge, J.~J.~Thomson Avenue, Cambridge CB3 0HE,
United Kingdom}

\begin{abstract}
  Quantum Monte Carlo calculations of the first-row atoms Li--Ne and
  their singly-positively-charged ions are reported.
  Multi-determinant-Jastrow-backflow trial wave functions are used
  which recover more than 98\% of the correlation energy at the
  Variational Monte Carlo (VMC) level and more than 99\% of the 
  correlation energy at the Diffusion Monte Carlo (DMC) level for both 
  the atoms and ions.  We obtain the first ionization potentials to 
  chemical accuracy.  We also report scalar relativistic corrections to 
  the energies, mass-polarization terms, and one- and two-electron
  expectation values.
\end{abstract}



\maketitle

\section{Introduction}
\label{sec:intro}

Quantum Monte Carlo (QMC) methods can yield highly accurate energies
for correlated quantum systems.  QMC calculations based on many-body
wave functions \cite{Foulkes2001} are considerably more accurate than
density functional theory (DFT) methods, and their accuracy rivals
that of the most sophisticated quantum chemistry methods.
The intrinsically parallel nature of QMC algorithms makes them
well-suited for taking advantage of the computing power offered by
modern massively-parallel machines.  The first-row atoms are a natural
set of systems to use in learning how to achieve chemical accuracy,
which is reached when an error of less than 1 kcal/mol $\simeq$ 1.6
m$E_h$ per atom $\simeq$ 43 meV per atom is achieved.  Accurate
benchmark data are available for light atoms, as are results from many
different electronic structure techniques.  The cost of all-electron
QMC calculations scales with the atomic number \cite{Ma2005} $Z$
roughly as $Z^{5.5}$, so that pseudopotentials must be used for heavy
atoms, but it is perfectly possible to perform highly-accurate
all-electron calculations for atoms up to at least the ten-electron
neon atom.

Here we apply the variational Monte Carlo (VMC) and diffusion Monte
Carlo (DMC) methods to calculating the ground-state energies and other
properties of the atoms Li--Ne and their singly-positively-charged ions.  
VMC expectation values of operators such as the Hamiltonian are calculated
with an approximate many-body trial wave function, $\Psi$, and the integrals 
are evaluated using a Monte Carlo technique.  The functional form of $\Psi$ 
is chosen to contain a number of parameters 
whose values are obtained by stochastic optimization.  Higher
accuracy is achieved in the DMC method by evolving the wave function
in imaginary time so that it decays towards the ground-state, while
the fixed-node approximation is used to maintain the fermionic
symmetry.  Both the VMC and DMC methods are variational, which is
helpful in monitoring the accuracy of the calculations and in
promoting cancellation of errors in energy differences.  The DMC
energy is bounded from above by the VMC energy and from below by the
exact energy.  These methods are discussed extensively in the
literature and we direct the reader to Refs.\
\onlinecite{Foulkes2001}, \onlinecite{Needs2010} and
\onlinecite{Nightingale1999} for a detailed description.

Our trial wave functions consist of a multi-determinant expansion
which describes near-degeneracy or static correlation, a Jastrow
factor which captures dynamic correlation, and a backflow
transformation which allows further variations in the nodal surface.
We recover over 98\% of the correlation energy for all the atoms and
ions studied at the VMC level and over 99\% at the DMC level.
Chemically-accurate values of the first ionization potentials are
obtained.  Total energies, scalar relativistic corrections to the
energies, mass-polarization terms, and one- and two-electron
expectation values are evaluated.  All of our QMC calculations were
performed using the \textsc{casino} package \cite{Needs2010}.

\section{Trial Wave Functions}
\label{sec:wfnc}

The all-electron multi-determinant-Jastrow-backflow wave functions
take the form:
\begin{equation}
\label{eq:wfnc}
\Psi(\mathbf{R})=e^{J(\mathbf{R};\mathbf{a})} \sum_{j=1}^{N_{\rm CSF}} 
c_j \sum_{k=1}^{N_{\rm det}^j} d_{k,j} D_{k,j}^\uparrow(\mathbf{x}_1,\ldots,
\mathbf{x}_{N_\uparrow}) D_{k,j}^\downarrow(\mathbf{x}_{N_{\uparrow+1}},
\ldots,\mathbf{x}_N),
\end{equation}
where $\bf{R}$ is the vector of electron positions, $J(\bf{R};\bf{a})$
is the Jastrow factor, and $D_{k,j}^\sigma(\bf{X})$ are the Slater 
determinants whose orbitals are evaluated at the backflow-transformed 
coordinates ${\bf x}_i={\bf r}_i+\mathbf{\xi}_i(\bf{R};\bf{b})$.  
$N_{\rm CSF}$ denotes the total number of configuration state functions 
(CSFs) and $N_{\rm det}^j$ is the number of determinants in the $j$th CSF.  
The vector $\bf{a}$ denotes the parameters in the Jastrow factor, $\bf{b}$
those in the backflow transformation, and 
${\bf c}=(c_1,\ldots,c_{N_{\rm CSF}})$ the coefficients of the CSFs. 
The coefficients of the determinants 
${\bf d}=(d_{1,1},\ldots,d_{N_{\rm det},N_{\rm CSF}})$ are held fixed 
to maintain the proper symmetry of the CSFs.

The Slater determinants and CSFs were generated using the atomic
multi-configuration Hartree-Fock (MCHF) package \textsc{atsp2k}
\cite{FroeseFischer2007}.  We allowed single- and double-excitations
from the HF ground-state configuration defined by the Aufbau principle
up to configurations with principal quantum number $n\leq7$ and
orbital angular momentum quantum number $l\leq4$.  Terms representing
excitations from the $1s^2$ core were used for Li, Li$^+$ and Be$^+$
to ensure that double-excitations were included.  The CSFs with the
largest weights were included in $\Psi$.  Core excitations
significantly lowered the MCHF energy of the Be atom, but they did not
improve the VMC energy and were therefore not included in the QMC
calculations.  Excitations from the core become less important for
larger $Z$, and we did not include them for systems with more than 
three electrons.  The high-energy excited-state configurations in 
the MCHF expansion mostly describe electron-electron cusps, which are 
captured by the Jastrow factor in QMC calculations.  The high-energy 
MCHF excitations are therefore expected to be much less important in 
the QMC calculations than in the MCHF ones.  Indeed, including
very-high-energy excitations serves only to hinder the optimization
procedure (see below) and worsen $\Psi$.  We tested wave functions
containing 1, 20 and 50 CSFs for all the atoms, and finally used 50
CSFs for all atoms and ions except O, O$^+$, F and F$^+$, for which we
used 100 CSFs. The number of determinants ranged from 171 (Li$^+$) to
4613 (F$^+$).

We used a modified form \cite{L'opezR'ios2010} of the polynomial
Jastrow factor proposed by Drummond \textit{et al.}\
\cite{Drummond2004}, consisting of an expansion in powers of
$r/(r^\beta+\alpha)$, where $r$ is the inter-particle separation and
$\alpha$ and $\beta$ are optimizable parameters, with $\beta$
constrained to be greater than unity.  The optimal values of $\alpha$
were found to lie within the range 0.5--17.1 and those of $\beta$ within
the range 1.05--4.67 for the atoms and ions studied.  This
modification removes the need for cut-offs at large inter-particle
separations, as the basis functions decay to zero at large $r$.  Based
on our tests, we chose expansion orders of 8
for the electron-electron and electron-nucleus parts of the Jastrow
factor and an expansion order of 4 for the terms in the
electron-electron-nucleus Jastrow factor, 
which gave a total of 118 optimizable parameters.

The backflow transformation of L\'opez R\'ios \textit{et al.}\
\cite{L'opezR'ios2006} was used, with electron-electron and 
electron-nucleus functions of expansion order 8
and an electron-electron-nucleus function of expansion order 3,
resulting in a further 142 optimizable parameters.  When using a backflow 
transformation, each orbital must be evaluated at each electron 
configuration, which significantly increases the computational cost.  
We have found it to be much more efficient to move the electrons 
individually in QMC calculations, even with backflow wave functions 
\cite{L'opezR'ios2006,Lee2010}, because the correlation length is shorter.

We used identical parameter values for pairs of up-spin electrons and
pairs of down-spin electrons in both the Jastrow and backflow functions.  
This significantly reduced the number of variable parameters without any
noticeable loss in wave function quality.  The parameter values for
the anti-parallel-spin channel were allowed to differ from those of
the parallel-spin channel.  The parallel and anti-parallel-spin cusp
conditions were imposed in the Jastrow factor, and the backflow
transformation does not introduce cusps \cite{L'opezR'ios2006}.

\section{Optimization}
\label{sec:opt}

Various stochastic methods have been developed for optimizing
many-body wave functions in QMC calculations
\cite{Umrigar1988,Kent1999,Drummond2005,Umrigar2007,Toulouse2007}.
Methods based on minimizing the variance of the local energies
obtained from a VMC calculation can be robust and effective
\cite{Umrigar1988,Kent1999,Drummond2005}, as is the related technique
of minimizing the mean absolute deviation of the local energies from
the median local energy (MAD minimization)\cite{Needs2010}.  
Simple implementations of variance minimization are very poor 
at optimizing nodal surfaces.  The reason for this is that the particle 
configurations are fixed within an optimization cycle, but changing 
parameters which alter the nodal surface may move it through the fixed 
configurations. The local energy diverges when the nodal surface 
coincides with configurations, leading to a poor optimization.
Variance-based optimization schemes can be effective in optimizing
nodal surfaces if the values of the local energies and/or the weights
of configurations near the nodal surface are limited
\cite{Drummond2005}.  However, lower total energies can be achieved by
minimizing the VMC energy itself \cite{Umrigar2007,Toulouse2007}.  The
energy minimization scheme of Umrigar and coworkers
\cite{Umrigar2007,Toulouse2007} is quite robust and is also extremely
effective in optimizing linear parameters in $\Psi$ such as CSF
coefficients.  We found MAD minimization to be superior to energy
minimization for optimizing the cut-off functions, and superior to
variance minimization methods for optimizing parameters which alter
the nodal surface.  We have therefore used MAD minimization in the
early stages of the optimizations, but the final optimizations are
performed with energy minimization.

We tested optimization of the single-particle orbitals for N, O and F,
but found this to have a negligible effect, in agreement with 
previous atomic studies \cite{Brown2007,Drummond2006}.

We tested several optimization schemes that could potentially reduce 
the computational effort of wave function optimization.  The Jastrow 
factor and backflow transformation were optimized for a single 
determinant and then applied to the multi-determinant expansion of a 
B wave function containing 50 CSFs.  Optimizing the CSF coefficients 
while holding the Jastrow factor and backflow parameters fixed 
improved the wave function but the final results remained unsatisfactory.  
This may be expected as the Jastrow and backflow functions attempt to 
compensate for some of the dynamical correlation which is then included 
via the CSFs.  The CSF coefficients in the B wave function were optimized 
for one final cycle, as energy minimization of linear coefficients is in 
general very robust.  No improvement was observed, confirming that energy
minimization is able to optimize the linear and non-linear
parameters simultaneously.

As the optimization process is currently the most costly step in
human time and consumes a substantial fraction of the computer time,
it is desirable to establish an optimization strategy which is
reliable for all of the atoms and ions and may be useful in other
systems.  Of the several optimization strategies tried, the following
consistently gave the best results and was used for all of the final
results reported here:
\begin{enumerate}
 \item Set the CSF coefficients to their MCHF values and the Jastrow 
  parameters $\bf{a}$ to zero. Note that the Jastrow factor is non-zero 
  as the term enforcing the cusp condition is still present.
 \item Generate a set of VMC configurations\footnote{The first set of VMC 
  configurations are drawn from the MCHF wave function.} and optimize the 
  Jastrow parameters $\bf{a}$ including the Jastrow  basis function 
  parameters $\alpha$ and $\beta$, and the CSF coefficients $\bf{c}$, using 
  MAD minimization. We refer to this step as an `optimization cycle'.  
 \item Run two more optimization cycles using the parameters obtained in 
  the previous cycle as initial parameters.
 \item Optimize the wave function parameters $\bf{a}$ and $\bf{c}$
  using VMC energy minimization until converged (usually about 5--8 
  cycles).  The Jastrow basis function parameters $\alpha$ and $\beta$ 
  are not re-optimized at this stage.
 \item Introduce backflow functions with the parameters $\bf{b}$
  initially set to zero, and optimize all wave function parameters
  ($\bf{a}$, $\bf{b}$, $\bf{c}$), including $\alpha$ and $\beta$, and
  the backflow cut-off parameters, using MAD minimization until
  converged (usually about 3 cycles).
 \item Use VMC energy minimization to optimize wave function parameters
  ($\bf{a}$, $\bf{b}$, $\bf{c}$) until converged (usually about 5--8
  cycles).  The Jastrow basis function parameters and backflow cut-off
  parameters are not re-optimized.
\end{enumerate}

\begin{figure}[h]
  \begin{center}
    \includegraphics[width=1.00\textwidth]{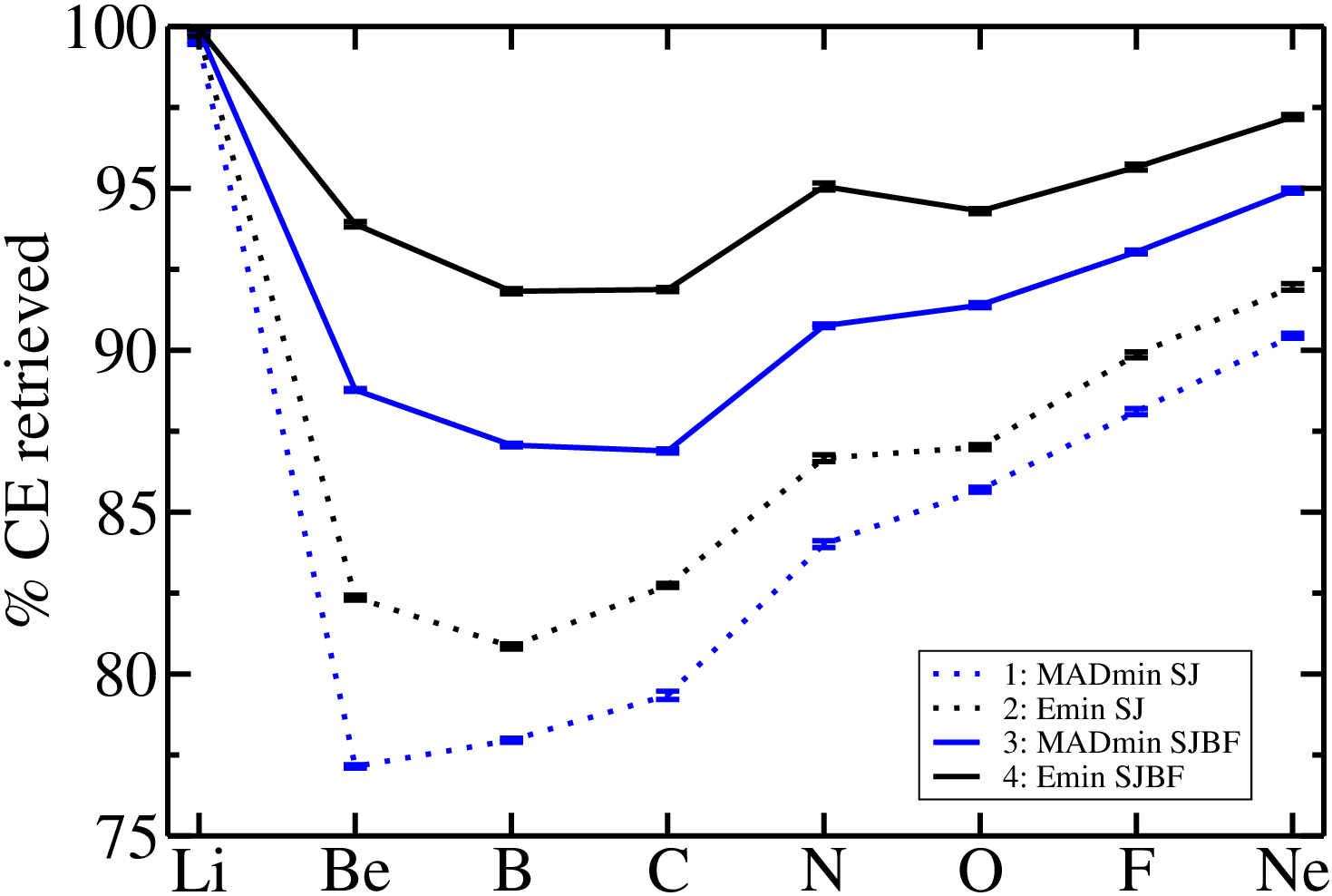}
    \caption{Percentages of the correlation energy (\% CE) retrieved for 
      single-determinant Slater-Jastrow (SJ) and Slater-Jastrow-backflow 
      (SJBF) wave functions using mean absolute deviation minimization 
      (MADmin) and energy minimization (Emin).
      \label{fig:stages_opt}}
  \end{center}
\end{figure}

The improvements in the VMC energies of the atoms at different levels
of optimization are shown in Fig.\ \ref{fig:stages_opt}.  The figure
clearly shows that the VMC energy minimization scheme yields
significantly larger percentages of the correlation energy than MAD
minimization.  While this strategy has not been tested for any other
systems, we expect it to work well in many cases.

\section{Results and Discussion}
\label{sec:results}

\subsection{Atomic and ionic energies}
\label{subsec:energy}

The VMC optimizations were performed using $5\times10^4$
statistically-independent particle configurations. One measure of wave 
function quality is its variance, and Table \ref{tab:var} reports the 
variances of optimized Slater-Jastrow and Slater-Jastrow-backflow wave 
functions for both single-determinant and multi-determinant Slater forms.  
The variance is reduced by approximately a factor of 2 or more when a 
multi-determinant expansion is introduced to a Slater-Jastrow-backflow 
wave function.  The improvement is particularly large for Be, B and C, 
where the variance drops by factors of approximately 10, 8 and 4, 
respectively. This reflects the strong 2s--2p near-degeneracy effects 
exhibited in these systems and determinants beyond the HF ground-state 
configuration must be included to capture the static correlation.  
With a few exceptions, the variance decreases as the variational
energy decreases for a given system.  The variance of a wave function 
after MAD minimization also approaches that after energy minimization as 
the variational energy decreases.
 
\begingroup
\begin{table}[h]
  \caption{VMC variances for single-determinant (SD) and 
    multi-determinant (MD) Slater-Jastrow (SJ) and 
    Slater-Jastrow-backflow (SJBF) wave functions.  
    All variances are in atomic units and the numbers in parentheses 
    indicate the statistical uncertainty in the last digit shown.
    \label{tab:var}}
\begin{tabular}{lr@{.}lr@{.}lr@{.}lr@{.}l}
\hline
& \multicolumn{2}{c}{SD-SJ} & \multicolumn{2}{c}{SD-SJBF} &
  \multicolumn{2}{c}{MD-SJ} & \multicolumn{2}{c}{MD-SJBF} \\
 \hline
 \hline
Li &$ 0$&$00274(3) $&$ 0$&$00130(1) $&$ 0$&$00193(5) $&$ 0$&$00067(2) $\\
Be &$ 0$&$0443(2)  $&$ 0$&$0524(6)  $&$ 0$&$01066(6) $&$ 0$&$00526(4) $\\
B  &$ 0$&$0915(2)  $&$ 0$&$1434(8)  $&$ 0$&$0326(2)  $&$ 0$&$01867(8) $\\
C  &$ 0$&$1941(8)  $&$ 0$&$1784(7)  $&$ 0$&$0819(5)  $&$ 0$&$0473(5)  $\\
N  &$ 0$&$340(1)   $&$ 0$&$263(1)   $&$ 0$&$2198(5)  $&$ 0$&$1126(5)  $\\
O  &$ 0$&$548(1)   $&$ 0$&$4763(9)  $&$ 0$&$442(1)   $&$ 0$&$353(1)   $\\
F  &$ 0$&$846(3)   $&$ 0$&$619(2)   $&$ 0$&$644(3)   $&$ 0$&$493(1)   $\\
Ne &$ 1$&$233(3)   $&$ 0$&$797(7)   $&$ 0$&$623(7)   $&$ 0$&$361(2)   $\\
 \hline
 \hline
\end{tabular}
\end{table}
\endgroup

The DMC calculations were performed with a target population of 2048 
DMC configurations and a minimum of $10^5$ steps and a time-step
corresponding to the smaller of the two used in Ref.\
\onlinecite{Brown2007}, ranging from 0.00375 a.u.\ for Li to 0.00070
a.u.\ for Ne.  These calculations \cite{Brown2007} already showed that
the errors from this choice of time steps is negligible, and the
corresponding errors in the current calculations should be even
smaller as the trial wave functions are superior.

\begin{figure}[h]
  \centering
    \includegraphics[width=1.00\textwidth]{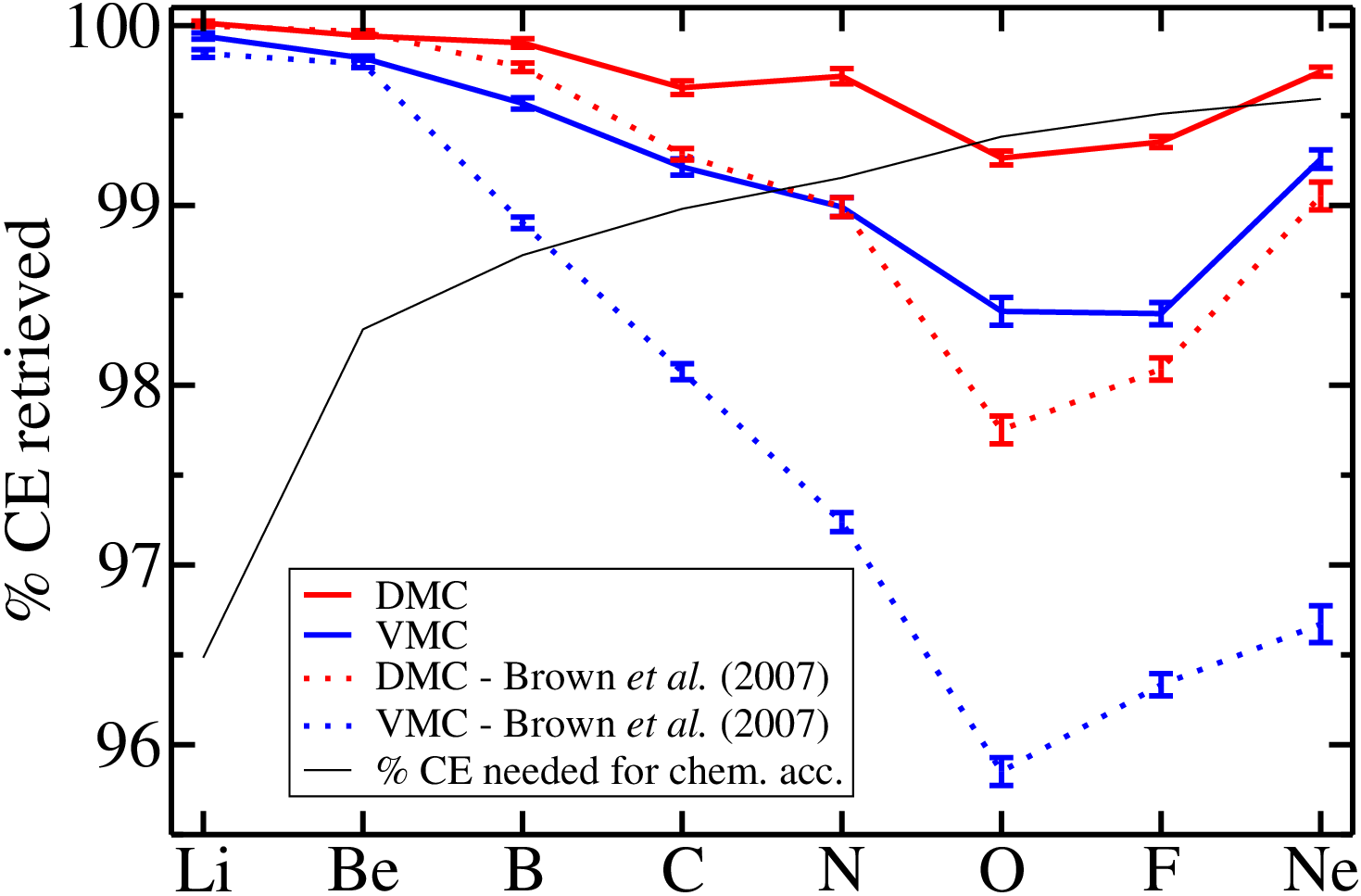}
    \caption{Percentages of the correlation energy (\% CE) retrieved for 
      each atom within VMC and DMC.  Chemical accuracy is achieved for 
      Li--N and Ne at the DMC level.
      \label{fig:atom_ce}}
\end{figure}

\begin{figure}[h]
  \centering
    \includegraphics[width=1.00\textwidth]{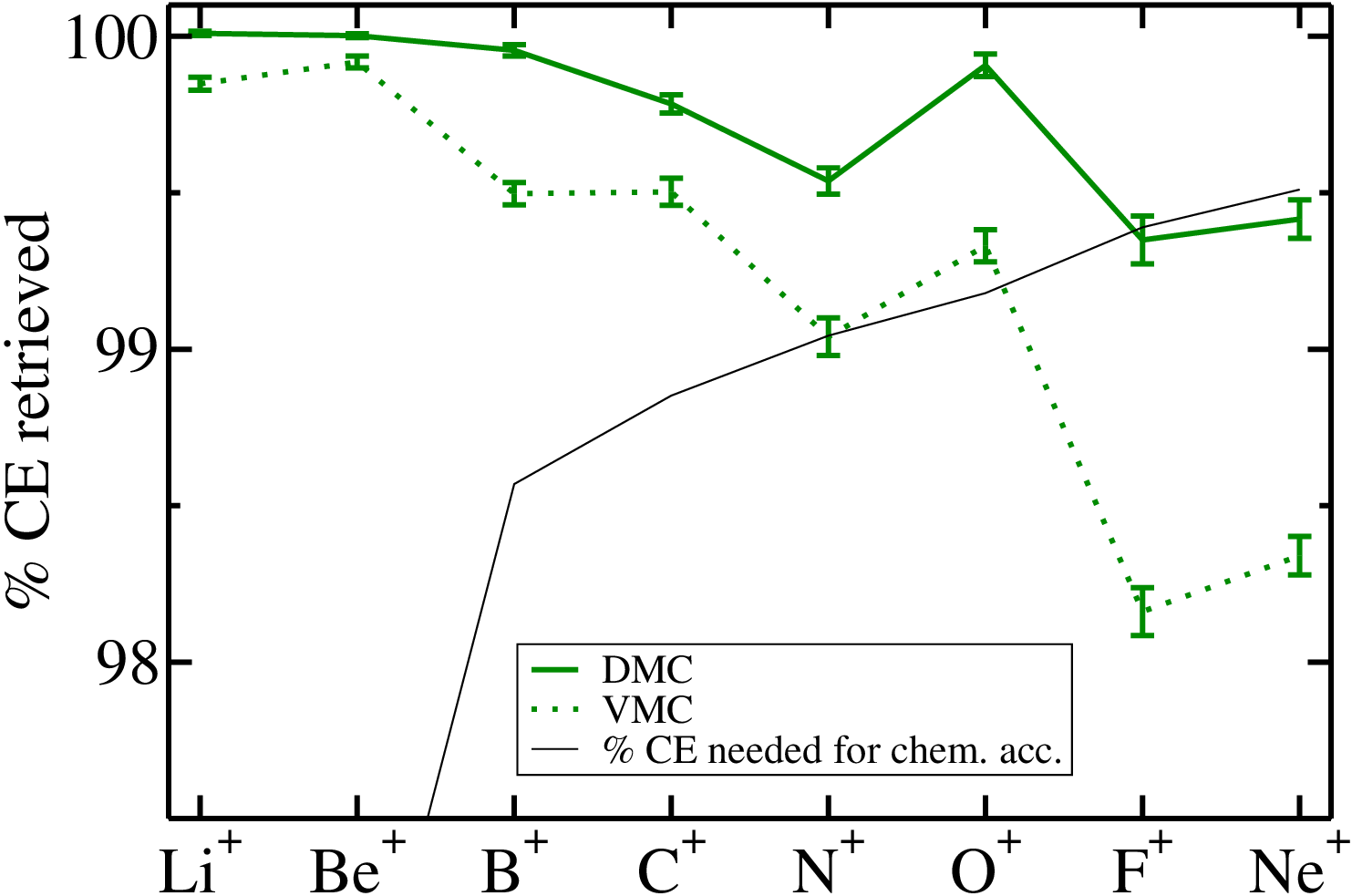}
    \caption{Percentages of the correlation energy (\% CE) retrieved for 
      each ion within VMC and DMC.  Chemical accuracy is achieved for
      Li$^+$--O$^+$ at the DMC level. The values for F$^+$ and Ne$^+$ 
      are within statistical uncertainty of chemical accuracy.
      \label{fig:ion_ce}}
\end{figure}

Table \ref{tab:pce} gives the VMC and DMC energies and percentages of
the correlation energy retrieved for each of the atoms and ions
studied.  The reference non-relativistic energies, assuming a clamped
point nucleus, are taken from Refs.\
\onlinecite{Puchalski2008,Chakravorty1993}. Percentages of the
correlation energy retrieved at the VMC and DMC levels for the neutral 
atoms in the present work and those of Ref.\ \onlinecite{Brown2007} are 
compared in Fig.\ \ref{fig:atom_ce} and for singly-charged ions are shown 
in Fig.\ \ref{fig:ion_ce}.  In both figures, the percentage of the 
correlation energy required to achieve chemical accuracy is indicated.

\begingroup
\squeezetable
\begin{table}[h]
  \caption{VMC and DMC energies of the first-row atoms and ions.  Also 
    included are Hartree-Fock energies $E_{\rm HF}$ calculated using 
    \textsc{atsp2k} \cite{FroeseFischer2007}, the reference energies 
    $E_{\rm ref}$ \cite{Puchalski2008,Chakravorty1993}, the correlation 
    energies $E_{\rm HF}-E_{\rm ref}$, and the percentage of the correlation 
    energy recovered at the VMC level (VMC-corr\%) and DMC level (DMC-corr\%). 
    All energies are in atomic units and the numbers in parentheses indicate 
    the statistical uncertainty in the last digit shown.
    \label{tab:pce}}
\begin{tabular}{lr@{.}lr@{.}lr@{.}lr@{.}lr@{.}lr@{.}lr@{.}lr@{.}l}
\hline
& \multicolumn{2}{c}{Li ($^{2}$S)} & \multicolumn{2}{c}{Be ($^{1}$S)} &
\multicolumn{2}{c}{B ($^{2}$P)} & \multicolumn{2}{c}{C ($^{3}$P)} & 
\multicolumn{2}{c}{N ($^{4}$S)} & \multicolumn{2}{c}{O ($^{3}$P)} & 
\multicolumn{2}{c}{F ($^{2}$P)} & \multicolumn{2}{c}{Ne ($^{1}$S)} \\ 
\hline
\hline
 VMC &
$ -7$&$478034(8)$ & $ -14$&$66719(1)$ & $ -24$&$65337(4)$ & $ -37$&$84377(7)$ &
$ -54$&$5873(1) $ & $ -75$&$0632(2) $ & $ -99$&$7287(2) $ & $-128$&$9347(2)$ \\ 

 DMC &
$ -7$&$478067(5)$ & $ -14$&$667306(7)$ & $ -24$&$65379(3)$ & $ -37$&$84446(6)$ &
$ -54$&$58867(8)$ & $ -75$&$0654(1)$ & $ -99$&$7318(1)$ & $ -128$&$9366(1)$ \\

 $E_{\mathrm{HF}}$ & 
$  -7$&$432727$ & $ -14$&$573023$ & $ -24$&$529061$ & $ -37$&$688619$ &
$ -54$&$400934$ & $ -74$&$809398$ & $ -99$&$409349$ & $-128$&$547098$ \\

 $E_{\mathrm{ref}}$&
$  -7$&$47806032$ & $ -14$&$66736$ & $ -24$&$65391$ & $ -37$&$8450$ &
$ -54$&$5892$ & $ -75$&$0673$ & $ -99$&$7339$ & $-128$&$9376$ \\

 $E_{\mathrm{HF}}-E_{\mathrm{ref}}$ &
$ 0$&$0453333 $ & $ 0$&$094337  $ & $ 0$&$124849  $ & $ 0$&$156381  $ &
$ 0$&$188266  $ & $ 0$&$257902  $ & $ 0$&$324551  $ & $ 0$&$390502  $ \\

 VMC-corr\% &
$99$&$94(2)\%$ & $99$&$82(1)\%$ & $99$&$57(3)\%$ & $99$&$21(4)\%$ &           
$98$&$99(5)\%$ & $98$&$41(8)\%$ & $98$&$40(6)\%$ & $99$&$26(5)\%$ \\

 DMC-corr\% &
$100$&$01(1)\%$ & $ 99$&$943(7)\%$ & $ 99$&$90(2)\%$ & $ 99$&$65(4)\%$ &
$ 99$&$72(4)\%$ & $ 99$&$26(4)\%$ & $ 99$&$35(3)\%$ & $ 99$&$74(3)\%$ \\
 \hline
 \hline
& \multicolumn{2}{c}{Li$^+$ ($^{1}$S)} & \multicolumn{2}{c}{Be$^+$ ($^{2}$S)} &
\multicolumn{2}{c}{B$^+$ ($^{1}$S)} & \multicolumn{2}{c}{C$^+$ ($^{2}$P)} & 
\multicolumn{2}{c}{N$^+$ ($^{3}$P)} & \multicolumn{2}{c}{O$^+$ ($^{4}$S)} & 
\multicolumn{2}{c}{F$^+$ ($^{3}$P)} & \multicolumn{2}{c}{Ne$^+$ ($^{2}$P)} \\ 
 \hline
 \hline
 VMC &
$ -7$&$279844(9)$ & $ -14$&$324721(9)$ & $ -24$&$34836(4)$ & $ -37$&$43034(6)$ &
$ -54$&$0530(1) $ & $ -74$&$5655(1) $ & $ -99$&$0880(2) $ & $-128$&$1377(2) $ \\

 DMC &
$-7$&$279914(3)$ & $ -14$&$324761(3)$ & $ -24$&$34887(2) $ & $ -37$&$43073(4)$ &
$ -54$&$05383(7) $ & $ -74$&$56662(7)$ & $ -99$&$0911(2) $ & $-128$&$1412(2)$ \\

 $E_{\mathrm{HF}}$ &
$  -7$&$236415$ & $ -14$&$277395$ & $ -24$&$237575$ & $ -37$&$292224$ &
$ -53$&$888005$ & $ -74$&$372606$ & $ -98$&$831720$ & $-127$&$817814$ \\

 $E_{\mathrm{ref}}$&
$  -7$&$27991$ & $ -14$&$32476$ & $ -24$&$34892$ & $ -37$&$43103$ &
$ -54$&$0546$ & $ -74$&$5668$ & $ -99$&$0928$ & $-128$&$1431$ \\

 $E_{\mathrm{HF}}-E_{\mathrm{ref}}$ &
$ 0$&$043495 $ & $ 0$&$047365 $ & $ 0$&$111345 $ & $ 0$&$138806 $ &
$ 0$&$166595 $ & $ 0$&$194194 $ & $ 0$&$26108 $ & $ 0$&$325286 $ \\

 VMC-corr\% &
$99$&$85(2)\%$ & $99$&$92(2)\%$ & $99$&$50(4)\%$ & $99$&$50(4)\%$ & 
$99$&$04(6)\%$ & $99$&$33(5)\%$ & $98$&$16(8)\%$ & $98$&$34(6)\%$ \\ 

 DMC-corr\% &
$100$&$009(7)\%$ & $100$&$002(6)\%$ & $ 99$&$96(2)\%$ & $ 99$&$78(3)\%$ &
$ 99$&$54(4)\%$ & $ 99$&$91(4)\%$ & $ 99$&$35(8)\%$ & $ 99$&$42(6)\%$ \\
 \hline
 \hline
\end{tabular}
\end{table}
\endgroup

There are several differences between the wave functions used in the
present study and Ref.\ \onlinecite{Brown2007}.  While both
calculations relied on the energy minimization scheme of Refs.\
\onlinecite{Umrigar2007,Toulouse2007}, our current implementation is
more effective and robust.  For example, Brown \textit{et al.}\
\cite{Brown2007} were unable to lower the VMC energy of Ne using a
multi-determinant expansion, which was easily achieved in the present
study.  The present optimization strategy is significantly different
as we use MAD minimization to first optimize the non-linear parameters
at each stage.  Brown \textit{et al.}\ \cite{Brown2007} used a Jastrow
factor based on an expansion in $r$, while we have used an expansion
in powers of $r/(r^\beta+\alpha)$.  We have also employed a larger
number of CSFs.

We have obtained more than 99\% of the correlation energy at the DMC 
level for all of the atoms and ions, and at the VMC level for all atoms
except O and F and all ions except F$^{+}$ and Ne$^{+}$.  This is a
substantially higher accuracy than has been achieved in the
all-electron QMC calculations reported in the literature
\cite{Brown2007,Buend'ia2009,Casula2003,Hongo2006}.  For example, the
lowest percentage of the correlation energy achieved for a neutral atom 
in the present study at the VMC level is 98.40(6)\% for F, whereas the
best previous VMC calculation gave 96.33(6)\%, and our lowest
percentage in DMC is 99.26(4)\% for O compared with the best previous
value of 97.83(8)\% \cite{Brown2007}.  We calculated the virial ratios
for the atoms and ions in both VMC and DMC, finding them to be within
one standard error of the exact value of 2 in each case, with the
standard errors lying within the range 0.001--0.02.

\subsection{Ionization potentials}
\label{subsec:ip}

\begin{figure}[h]
  \centering
    \includegraphics[width=1.00\textwidth]{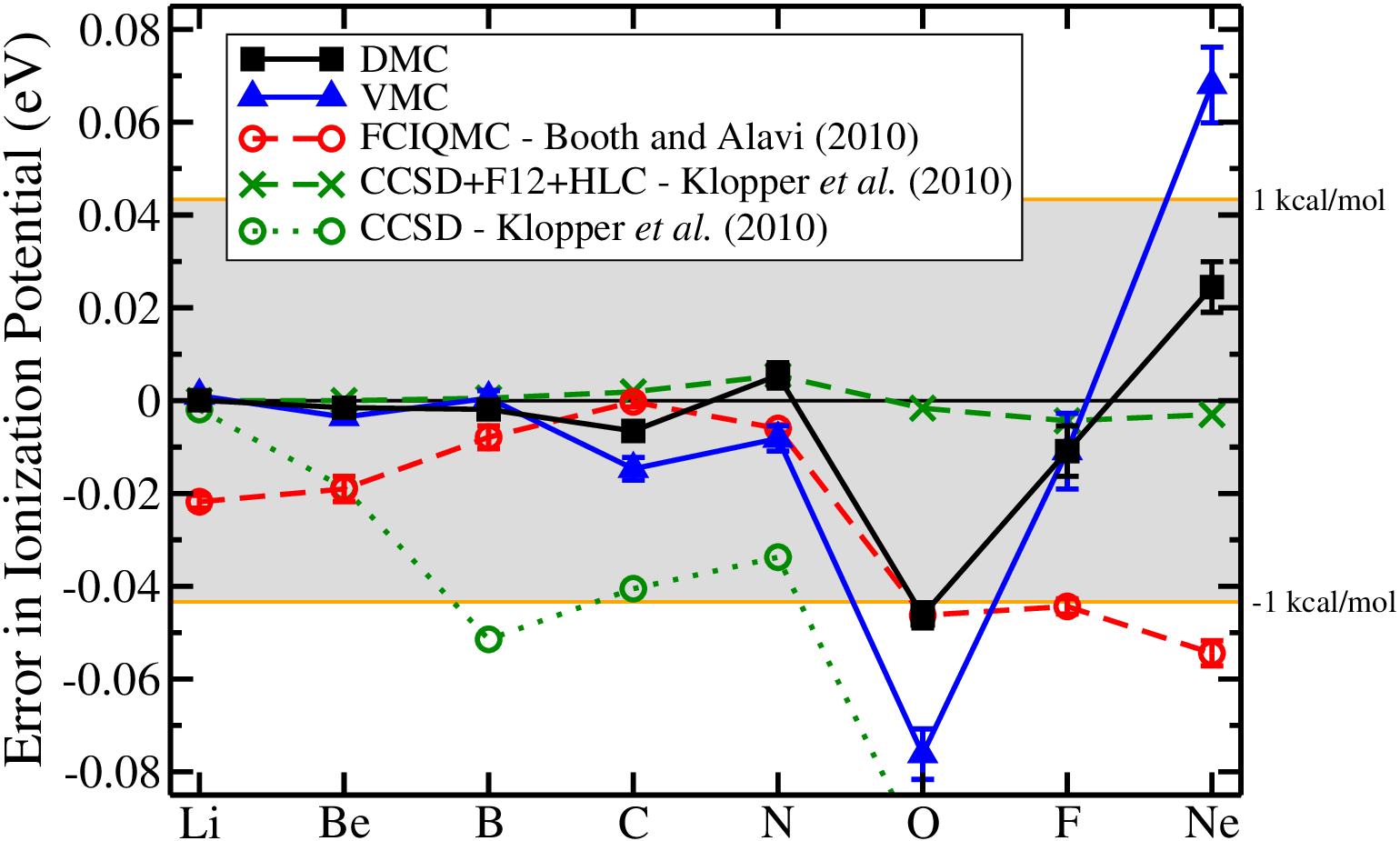}
    \caption{Errors in the ionization potentials ($\Delta=$ IP$_{\rm
        calc}-$IP$_{\rm ref}$) for the first-row atoms obtained at the
      VMC and DMC levels compared to those from FCI-QMC, CCSD and 
      CCSD-F12-HLC.  The reference values are taken from
      Ref.\ \onlinecite{Chakravorty1993}.  The shaded region
      represents chemical accuracy.
      \label{fig:ip}}
\end{figure}

Although the total atomic energies can be measured as the sum of the
ionization energies, they are not quantities of significant chemical
interest.  In quantum chemistry one is normally interested in energy
differences for which the cancellation of errors between calculations
is important.  We have therefore calculated the first ionization
potentials (IPs) of the atoms Li--Ne as energy differences between the
neutral and singly-ionized states.  The errors in the calculated IPs
from those computed using values from Ref.\
\onlinecite{Chakravorty1993} are shown in Fig.\ \ref{fig:ip}.  Data
from the stochastic full configuration interaction method \cite{Booth2010} 
(FCI-QMC) with an aug-cc-pVQZ basis set for Li, Be and Ne and
an aug-cc-pV5Z basis for B--F are shown, together with coupled cluster
single and double excitation (CCSD) data with a d-aug-cc-pwCV5Z basis
and CCSD-F12-HLC data \cite{Klopper2010}.
Each CCSD-F12-HLC energy is the sum of the CCSD energy, an F12
energy which corrects for the finite basis set, and a higher-level
correction (HLC) which accounts for the treatment of excitations
beyond the doubles in CCSD.  It is likely that the CCSD-F12-HLC
results \cite{Klopper2010} are even more accurate than the data of
Ref.\ \onlinecite{Chakravorty1993} that we have used as a reference,
as they obtain results in closer agreement with experiment when
corrections for relativistic effects and the finite nuclear mass are
included.  However, Klopper \textit{et al.}\ \cite{Klopper2010} did
not give values for the total energies of the atoms, and therefore we
have used the data of Ref.\ \onlinecite{Chakravorty1993} to avoid
using different reference data for the total energies and IPs.  The
differences from using the IP data of Klopper \textit{et al.}\
\cite{Klopper2010} are small, as can be seen in Fig.\ \ref{fig:ip}.
Using this data as the reference would not significantly affect the 
comparisons for Li and Be, but it would slightly worsen the agreement 
with our results for B, C and Ne and slightly improve it for N, O and F.

The IPs from DMC are within statistical error of chemical accuracy 
of the reference data for all atoms.  Our errors are smaller than or equal to
those in the CCSD results with a d-aug-cc-pwCV5Z basis
\cite{Klopper2010} for all atoms, and smaller than those of the FCI-QMC 
calculations of Booth and Alavi \cite{Booth2010} for all atoms except C.  
The mean deviation, mean absolute deviation and maximum deviation of 
the IPs from the reference values for these methods and those obtained 
in DFT using the B3LYP, LSDA and PBE-GGA density functionals are 
presented in Table \ref{tab:mae}.  The mean absolute deviations of the
DFT IPs are between 24 and 30 times larger than for our DMC calculations.

The FCI-QMC approach \cite{Booth2010} is exact up
to a basis set convergence error and a small statistical error.  They
consistently underestimate IPs, perhaps because there are fewer
electron-electron and electron-nucleus cusps in an ion than in the
corresponding neutral atom.  As is clear from Fig.\ \ref{fig:ip}, we
similarly underestimate the IPs in all cases except Ne, but for a
different reason, as explained below.

The critical approximation made in DMC is the fixed-node
approximation, which is necessary to evade the fermion sign problem.
In general, the nodal structure of an ion is easier to describe than 
that of the corresponding neutral atom.  However, for closed-shell atoms 
such as Ne, the initial restricted HF atomic nodal surface is superior 
to that of the open-shell ion.  The energy of the neutral atom is more 
accurate and consequently the IP is overestimated.

To summarize, the additional complexity that arises for larger system
sizes manifests itself differently in FCI and DMC. In the former,
describing the electron-electron cusps becomes more challenging and
requires a larger expansion in determinants, whereas the nodal
structure of the larger system is more difficult to describe in DMC.

\begin{table}[h]
  \caption{Comparison of the mean deviation ($\overline{\Delta}$), mean 
    absolute deviation ($\overline{\lvert\Delta\rvert}$) and maximum 
    deviation ($\Delta_{\rm max}$) of the ionization potentials obtained 
    from several electronic structure methods.  Deviations are from the 
    reference non-relativistic, clamped point nucleus values of Ref.\ 
    \onlinecite{Chakravorty1993}. Averages were taken over Li--Ne, unless 
    otherwise indicated.  All values are in electron volts and the numbers 
    in parentheses indicate the statistical uncertainty, if present, in the 
    last digit shown.
    \label{tab:mae}}
\begin{tabular}{lr@{.}lr@{.}lr@{.}l}
\hline
& \multicolumn{2}{c}{$\overline{\Delta}$} &
  \multicolumn{2}{c}{$\overline{\lvert\Delta\rvert}$} &
  \multicolumn{2}{c}{$\Delta_{\rm max}$} \\ 
\hline
\hline
VMC                          & $ -0$&$005(2)  $&$ 0$&$023(2)  $&$ 0$&$076(5) $\\ 
DMC                          & $ -0$&$005(1)  $&$ 0$&$012(1)  $&$ 0$&$046(2) $\\
FCI-QMC\footnotemark[1]      & $ -0$&$0250(7) $&$ 0$&$0250(7) $&$ 0$&$054(3) $\\
CCSD\footnotemark[2]         & $ -0$&$0586    $&$ 0$&$0585    $&$ 0$&$1140   $\\
CCSD-F12-HLC\footnotemark[2] & $ -0$&$0001    $&$ 0$&$0021    $&$ 0$&$0054   $\\
B3LYP\footnotemark[3]        & $  0$&$2925    $&$ 0$&$2924    $&$ 0$&$5206   $\\
LSDA\footnotemark[4]         & $  0$&$2657    $&$ 0$&$3521    $&$ 0$&$5447   $\\
PBE\footnotemark[4]          & $  0$&$1971    $&$ 0$&$2892    $&$ 0$&$4507   $\\
\hline
\hline
\end{tabular}
\footnotetext[1]{Ref.\ \onlinecite{Booth2010}.}
\footnotetext[2]{Ref.\ \onlinecite{Klopper2010}.}
\footnotetext[3]{Averages taken over B--Ne values. Ref.\ \onlinecite{DeProft1997}.}
\footnotetext[4]{Averages taken over Li--F values. Ref.\ \onlinecite{Ernzerhof1999}.}
\end{table}

\subsection{Other expectation values}
\label{subsec:expectation values}

Scalar relativistic and mass-polarization corrections to the energies
of the atoms and ions were computed using first-order perturbation
theory within VMC and DMC \cite{Kenny1995}. The mass-velocity,
one-electron Darwin, two-electron Darwin, retardation and
mass-polarization terms are given in Table \ref{tab:rel_corr}.  We did
not calculate the spin-orbit terms, although we note that a paper
reporting VMC results for spin-orbit energies of electronic states of
the He atom has recently appeared \cite{Alexander2010}.
As the VMC and DMC values agree to within one standard error in
most cases, we have decided not to give the extrapolated estimates
(2$\times$DMC-VMC) in Table \ref{tab:rel_corr} and instead we quote
only the DMC values.  Our results for Li, Li$^+$, Be and Be$^+$ are
close to those given in Refs.\ \onlinecite{King1998}, \onlinecite{Yan1998} 
and \onlinecite{Pekeris1962}, obtained from Hylleraas calculations,
and in Ref.\ \onlinecite{Pachucki2004}, from exponentially-correlated 
Gaussian calculations. 

\begingroup
\squeezetable
\begin{table}[h]
  \caption{Scalar relativistic terms: mass-velocity (MV), electron-nucleus 
    Darwin (D1), two-electron Darwin (D2), spin-spin contact interaction (SSC),
    retardation (Ret), and mass-polarization (MP) energies calculated at 
    the DMC level.  Values from the literature are given for Li, Li$^+$, Be 
    and Be$^+$.  All values are in atomic units and the numbers in 
    parentheses give the statistical uncertainty in the last digit shown.
    \label{tab:rel_corr}}
\begin{tabular}{lr@{.}lr@{.}lr@{.}lr@{.}lr@{.}l}
\hline
& \multicolumn{2}{c}{MV}& \multicolumn{2}{c}{D1}& \multicolumn{2}{c}{D2+SSC} & 
\multicolumn{2}{c}{Ret}& \multicolumn{2}{c}{MP} \\ 
\hline
\hline
 Li  &$ -0$&$00417(1)  $&$  0$&$00346(1)  $&$  0$&$0000914(3)  $&$  -0$&$0000232(1)  $&$   0$&$0000239(1)  $\\   
 Be  &$ -0$&$01439(2)  $&$  0$&$01181(2)  $&$  0$&$0002690(5)  $&$  -0$&$0000478(1)  $&$  0$&$00002815(9)  $\\   
 B   &$  -0$&$0368(1)  $&$   0$&$0300(1)  $&$   0$&$000598(2)  $&$  -0$&$0000585(7)  $&$   0$&$0000137(3)  $\\   
 C   &$  -0$&$0790(2)  $&$   0$&$0639(2)  $&$   0$&$001115(8)  $&$   -0$&$000017(2)  $&$  -0$&$0000178(5)  $\\   
 N   &$  -0$&$1504(6)  $&$   0$&$1207(7)  $&$    0$&$00185(4)  $&$    0$&$000147(5)  $&$   -0$&$000069(2)  $\\   
 O   &$  -0$&$2610(7)  $&$   0$&$2086(7)  $&$    0$&$00299(3)  $&$    0$&$000415(4)  $&$  -0$&$0001278(8)  $\\   
 F   &$   -0$&$424(1)  $&$    0$&$337(1)  $&$    0$&$00451(5)  $&$    0$&$000935(7)  $&$   -0$&$000195(1)  $\\   
 Ne  &$   -0$&$655(2)  $&$    0$&$518(2)  $&$    0$&$00646(8)  $&$     0$&$00186(1)  $&$   -0$&$000303(1)  $\\   
\hline
\hline

 Li$^+$  &$ -0$&$00411(1)  $&$  0$&$00341(1)  $&$  0$&$0000895(2)  $&$  -0$&$00002291(9)  $&$  0$&$00002298(8)  $\\   
 Be$^+$  &$ -0$&$01426(2)  $&$  0$&$01171(2)  $&$  0$&$0002649(3)  $&$   -0$&$0000485(1)  $&$  0$&$00002756(7)  $\\   
 B$^+$   &$  -0$&$0377(3)  $&$   0$&$0307(3)  $&$   0$&$000597(2)  $&$   -0$&$0000802(6)  $&$   0$&$0000305(3)  $\\   
 C$^+$   &$  -0$&$0796(3)  $&$   0$&$0643(3)  $&$   0$&$001115(5)  $&$    -0$&$000056(1)  $&$  -0$&$0000003(4)  $\\   
 N$^+$   &$  -0$&$1506(5)  $&$   0$&$1208(5)  $&$    0$&$00190(1)  $&$     0$&$000071(2)  $&$  -0$&$0000494(6)  $\\   
 O$^+$   &$   -0$&$263(2)  $&$    0$&$210(2)  $&$    0$&$00299(2)  $&$     0$&$000388(4)  $&$  -0$&$0001143(8)  $\\   
 F$^+$   &$   -0$&$425(2)  $&$    0$&$337(2)  $&$     0$&$0044(1)  $&$      0$&$00091(5)  $&$   -0$&$000183(3)  $\\   
 Ne$^+$  &$   -0$&$658(2)  $&$    0$&$519(2)  $&$    0$&$00636(8)  $&$     0$&$001719(9)  $&$   -0$&$000287(1)  $\\   
\hline
\hline

 Li     &$ -0$&$00418308\footnotemark[1]     $&$ 0$&$00347364\footnotemark[1]    $&$ 0$&$0000911359\footnotemark[1]    
        $&$ -0$&$0000232018\footnotemark[1]  $&$ 0$&$0000236819\footnotemark[2]  $\\
 Li$^+$ &$ -0$&$00413427\footnotemark[3]     $&$ 0$&$00343889\footnotemark[3]    $&$ 0$&$000089292\footnotemark[3]     
        $&$ -0$&$000022791\footnotemark[3]   $&$ 0$&$00002259816\footnotemark[3] $\\
 Be     &$ -0$&$01441539\footnotemark[4]     $&$ 0$&$011834014\footnotemark[4]   $&$ 0$&$0002685577\footnotemark[4] 
        $&$ -0$&$0000474909\footnotemark[4]  $&$ 0$&$0000278121\footnotemark[4]  $\\
 Be$^+$ &$ -0$&$0142882124\footnotemark[4]   $&$ 0$&$011745724\footnotemark[4]   $&$ 0$&$0002644146\footnotemark[4]    
        $&$ -0$&$00004845370\footnotemark[4] $&$ 0$&$0000273704\footnotemark[4]  $\\
\hline
\hline

\end{tabular}
\footnotetext[1]{Ref.\ \onlinecite{King1998}.}
\footnotetext[2]{Ref.\ \onlinecite{Yan1998}.}
\footnotetext[3]{Ref.\ \onlinecite{Pekeris1962}.}
\footnotetext[4]{Ref.\ \onlinecite{Pachucki2004}.}
\end{table}
\endgroup

In Table \ref{tab:exp_val_one_e} we report some one-electron
expectation values for the atoms and ions, while two-electron
expectation values are reported in Table \ref{tab:exp_val_two_e}.
Variational calculations using Hylleraas-type wave functions have
given very accurate results for three systems included in our
study: the Li atom \cite{King1995,Yan1995}, the Li$^{+}$ ion
\cite{Frolov2006}, and the Be$^{+}$ ion \cite{Frolov2009}.  Our
results for these systems are in good agreement with the Hylleraas
data.  The data available in the literature for systems with more than
three electrons are of much lower accuracy.  Cohen \textit{et al.}\
\cite{Cohen2004} have reported values of $\langle r_i^2 \rangle$ for
some atoms, including Be, B and C, calculated within unrestricted HF
theory and correlated theories such as FCI.  Electron correlation is
expected to reduce the size of atoms as measured by $\langle r_i^2
\rangle$.  Our value of $\langle r_i^2 \rangle$ for Be reported
in Table \ref{tab:exp_val_one_e} is slightly larger than the FCI
values of 16.27, while our values for B and C are slightly smaller 
than the FCI values of 15.54 (B) and 13.84 (C) \cite{Cohen2004}.  
In almost all cases the values of the one-electron expectation 
values (summed over the electrons) are larger for the neutral atoms 
than for the corresponding ions, as one would expect.  However, the
expectation values of $\langle \delta(r_i) \rangle$ and 
$\langle r_i^{-2} \rangle$, which are the most strongly weighted 
towards the region close to the nucleus, are larger for the ion than
for the neutral atom for B/B$^+$, and very similar for 
C/C$^+$--Ne/Ne$^+$.  The larger error bars and lower quality of the 
wave functions make it more difficult to draw firm conclusions for
C/C$^+$--Ne/Ne$^+$.

\begingroup
\squeezetable
\begin{table}[h]
  \caption{One-electron expectation values: electron moments 
    $\langle r_i^n \rangle$ for $-2 \leq n \leq 3$ and electron density at 
    the coalescence point $\langle \delta(r_i) \rangle$, summed over all 
    electrons $i$.  All values are in atomic units and the numbers in 
    parentheses indicate the statistical uncertainty in the last digit shown.
    \label{tab:exp_val_one_e}}
\begin{tabular}{lr@{\@}lr@{.}lr@{.}lr@{.}lr@{.}lr@{.}l}
\hline
& \multicolumn{2}{c}{$\langle \delta(r_i) \rangle$} & \multicolumn{2}{c}{$\langle r_i^{-2} \rangle$} 
& \multicolumn{2}{c}{$\langle r_i^{-1} \rangle$} & \multicolumn{2}{c}{$\langle r_i \rangle$} 
& \multicolumn{2}{c}{$\langle r_i^2 \rangle$} & \multicolumn{2}{c}{$\langle r_i^3 \rangle$} \\ 
\hline
\hline
 Li &$  13$&$.79(5) $&$  30$&$25(4) $&$   5$&$7193(4)  $&$  4$&$9842(3) $&$  18$&$300(2)  $&$   92$&$10(1) $\\
 Be &$  35$&$.30(6) $&$  57$&$59(3) $&$   8$&$4275(2)  $&$  5$&$9794(1) $&$ 16$&$2986(4)  $&$  57$&$078(2) $\\
 B  &$   71$&$.7(3) $&$  93$&$51(9) $&$  11$&$3993(7)  $&$  6$&$7446(3) $&$ 15$&$5322(9)  $&$  46$&$011(4) $\\
 C  &$  127$&$.2(4) $&$  138$&$8(1) $&$  14$&$7065(8)  $&$  7$&$1230(3) $&$ 13$&$7401(7)  $&$  33$&$940(3) $\\
 N  &$    206$&$(1) $&$  193$&$0(1) $&$  18$&$3491(9)  $&$  7$&$3612(2) $&$ 12$&$1750(5)  $&$  25$&$740(2) $\\
 O  &$    312$&$(1) $&$  257$&$1(2) $&$   22$&$271(1)  $&$  7$&$6364(2) $&$ 11$&$3283(5)  $&$  21$&$756(2) $\\
 F  &$    448$&$(2) $&$  330$&$8(2) $&$   26$&$537(1)  $&$  7$&$8166(2) $&$ 10$&$4132(4)  $&$  18$&$003(1) $\\
 Ne &$    619$&$(2) $&$  414$&$3(3) $&$   31$&$134(1)  $&$  7$&$9298(2) $&$  9$&$5220(4)  $&$ 14$&$8372(9) $\\
 \hline
 \hline
 Li$^+$ &$ 13$&$.60(4) $&$  29$&$81(4) $&$  5$&$3770(4) $&$ 1$&$14539(7) $&$ 0$&$89252(7) $&$  0$&$8830(1) $\\
 Be$^+$ &$ 35$&$.01(5) $&$  56$&$97(2) $&$  7$&$9760(2) $&$ 3$&$10220(6) $&$  6$&$5122(2) $&$ 18$&$7046(8) $\\
 B$^+$  &$  73$&$.5(8) $&$  93$&$94(9) $&$ 10$&$9332(7) $&$  4$&$1791(2) $&$  7$&$6318(5) $&$  17$&$736(2) $\\
 C$^+$  &$ 128$&$.1(6) $&$  138$&$9(1) $&$ 14$&$1589(8) $&$  4$&$9235(2) $&$  7$&$9284(4) $&$  16$&$072(1) $\\
 N$^+$  &$ 206$&$.3(9) $&$  193$&$2(2) $&$  17$&$727(1) $&$  5$&$4338(2) $&$  7$&$7161(4) $&$  13$&$742(1) $\\
 O$^+$  &$   314$&$(3) $&$  257$&$1(2) $&$  21$&$618(1) $&$  5$&$8097(2) $&$  7$&$3423(3) $&$ 11$&$6156(8) $\\
 F$^+$  &$   447$&$(3) $&$  331$&$3(2) $&$  25$&$811(1) $&$  6$&$1624(2) $&$  7$&$1305(3) $&$ 10$&$3950(7) $\\
 Ne$^+$ &$   621$&$(2) $&$  414$&$9(3) $&$  30$&$323(1) $&$  6$&$4236(2) $&$  6$&$7976(3) $&$  9$&$0808(5) $\\
 \hline
 \hline
\end{tabular}
\end{table}
\endgroup

\begingroup
\squeezetable
\begin{table}[h]
  \caption{Two-electron expectation values: inter-electronic moments 
    $\langle r_{ij}^n \rangle$ for $-2 \leq n \leq 3$, the electron-pair 
    density at the coalescence point $\langle \delta(r_{ij}) \rangle$
    and the mass-polarization term $-\langle \boldsymbol{\nabla}_i \cdot 
    \boldsymbol{\nabla}_j \rangle$, summed over all electron-pairs $ij$.
    All values are in atomic units and the numbers in parentheses indicate 
    the statistical uncertainty in the last digit shown.
    \label{tab:exp_val_two_e}}
\begin{tabular}{lr@{.}lr@{.}lr@{.}lr@{.}lr@{.}lr@{.}lr@{.}l}
\hline
& \multicolumn{2}{c}{$\langle \delta(r_{ij}) \rangle$} & \multicolumn{2}{c}{$\langle r_{ij}^{-2} \rangle$} 
& \multicolumn{2}{c}{$\langle r_{ij}^{-1} \rangle$} & \multicolumn{2}{c}{$\langle r_{ij} \rangle$} 
& \multicolumn{2}{c}{$\langle r_{ij}^2 \rangle$} & \multicolumn{2}{c}{$\langle r_{ij}^3 \rangle$} 
& \multicolumn{2}{c}{$-\langle \boldsymbol{\nabla}_i \cdot \boldsymbol{\nabla}_j \rangle$} \\ 
\hline
\hline
 Li &$ 0$&$546(2)  $&$   4$&$386(6) $&$  2$&$1993(1) $&$  8$&$6574(4) $&$  36$&$731(2) $&$  191$&$00(2) $&$   0$&$304(2) $\\
 Be &$ 1$&$608(3)  $&$   9$&$532(4) $&$ 4$&$37330(9) $&$ 15$&$2895(2) $&$ 52$&$9958(9) $&$ 223$&$436(6) $&$   0$&$466(1) $\\
 B  &$  3$&$57(1)  $&$   17$&$45(1) $&$  7$&$6657(3) $&$ 22$&$4696(6) $&$  67$&$189(2) $&$  244$&$63(1) $&$   0$&$272(6) $\\
 C  &$  6$&$66(5)  $&$   29$&$15(2) $&$ 12$&$5191(4) $&$ 29$&$0654(6) $&$  73$&$574(2) $&$ 225$&$629(9) $&$   -0$&$39(1) $\\
 N  &$  11$&$1(2)  $&$   45$&$74(2) $&$ 19$&$2241(5) $&$ 35$&$5092(6) $&$  77$&$550(2) $&$ 204$&$547(6) $&$   -1$&$78(5) $\\
 O  &$  17$&$9(2)  $&$   68$&$85(3) $&$ 27$&$9913(6) $&$ 42$&$5224(6) $&$  83$&$693(2) $&$ 200$&$810(6) $&$   -3$&$76(2) $\\
 F  &$  27$&$0(3)  $&$   99$&$76(4) $&$ 39$&$2235(7) $&$ 49$&$3275(7) $&$  87$&$469(2) $&$ 189$&$266(5) $&$   -6$&$82(4) $\\
 Ne &$  38$&$6(5)  $&$  140$&$37(5) $&$ 53$&$2639(8) $&$ 55$&$8986(6) $&$  89$&$587(1) $&$ 175$&$092(4) $&$  -11$&$23(5) $\\
 \hline
 \hline
 Li$^+$  &$   0$&$535(1) $&$   4$&$088(6) $&$  1$&$5684(1) $&$ 0$&$86210(7) $&$ 0$&$92684(9) $&$ 1$&$1888(2) $&$   0$&$293(1) $\\
 Be$^+$  &$   1$&$584(2) $&$   8$&$901(3) $&$ 3$&$24613(7) $&$ 5$&$26857(9) $&$ 13$&$0819(3) $&$ 39$&$457(1) $&$   0$&$456(1) $\\
 B$^+$   &$   3$&$57(1)  $&$   16$&$47(1) $&$  5$&$9706(3) $&$ 10$&$5411(3) $&$  24$&$698(1) $&$ 69$&$807(4) $&$   0$&$606(5) $\\
 C$^+$   &$   6$&$67(3)  $&$   27$&$57(2) $&$ 10$&$0797(4) $&$ 16$&$2100(4) $&$  34$&$253(1) $&$ 86$&$994(4) $&$  -0$&$007(9) $\\
 N$^+$   &$  11$&$33(7)  $&$   43$&$38(3) $&$ 15$&$9096(5) $&$ 21$&$9744(4) $&$  41$&$344(1) $&$ 93$&$139(3) $&$  -1$&$27(2)  $\\
 O$^+$   &$  17$&$8(1)  $&$    65$&$03(3) $&$ 23$&$7541(6) $&$ 27$&$8183(5) $&$  46$&$843(1) $&$ 94$&$164(3) $&$  -3$&$36(2)  $\\
 F$^+$   &$  26$&$3(9)  $&$    94$&$80(4) $&$ 33$&$9176(7) $&$ 34$&$0801(5) $&$  52$&$794(1) $&$ 98$&$132(3) $&$  -6$&$39(9)  $\\
 Ne$^+$  &$  38$&$0(5)  $&$   133$&$80(5) $&$ 46$&$7451(8) $&$ 40$&$2711(5) $&$  57$&$231(1) $&$ 97$&$664(2) $&$ -10$&$62(5)  $\\

 \hline
 \hline
\end{tabular}
\end{table}
\endgroup

\section{Conclusions}

We have calculated energies for the first-row atoms with
significantly more accuracy than previous DMC studies.  Our DMC
energies for the atoms heavier than Li and ions heavier than
Be$^+$ are the lowest so far reported from a variational method.  Our
DMC IPs are also superior to very recent FCI-QMC results
\cite{Booth2010}.  Our IPs are, however, substantially less accurate
than the CCSD-F12-HLC data of Klopper \textit{et al.}\
\cite{Klopper2010}.  Our DMC IPs are considerably better than the CCSD
values, but the addition of the F12 and HLC corrections leads to
errors which are roughly an order of magnitude smaller than in our DMC
calculations.  The DMC calculations have the feature that the
results are obtained from a single calculation, and the cost of
calculating the F12 and HLC corrections in the CCSD scheme will
increase very rapidly with the number of electrons.

It would be extremely useful if \textit{post hoc} corrections could be
developed for QMC methods.  One method which has shown some success in
VMC is to plot the total energy versus the variance of the local
energy using a set of trial wave functions of different qualities
\cite{Kwon1993}.  Such a plot normally shows an approximately linear
variation so that an extrapolation to zero variance can be performed.
The linear variation can be derived by assuming that the set of wave
functions differ by a term of the form $\epsilon \Phi({\bf R})$, where
$\epsilon$ is a parameter and $\Phi({\bf R})$ is an (unknown) wave
function, but there is no guarantee that this assumption is valid.
Perhaps a \textit{post hoc} correction scheme can be developed for DMC
calculations.

For the most difficult case of the O atom we obtained an error in the
energy of 2.17(8)\% in our 2007 study \cite{Brown2007} compared with
0.74(4)\% in the present study.  There is every prospect of making
substantial further improvements to the VMC and DMC results.  The
stochastic optimization techniques used to obtain the optimal values
of the wave function parameters have improved greatly in recent years,
mainly due to the work of Umrigar and collaborators
\cite{Umrigar2007,Toulouse2007}.  The development of VMC sampling 
strategies which allow more reliable and efficient optimization of 
wave functions is extremely promising \cite{Trail2010}.  There have 
been major improvements in the available wave function forms
\cite{L'opezR'ios2006,Bajdich2006,Bajdich2008,Marchi2009}, and many
more such developments can be expected in the coming years.  The cost
of the QMC calculations reported here increases rapidly with system 
size because of the use of a multi-determinant expansion.  However, 
we expect that the computational cost could be substantially reduced 
by using a more efficient representation such as geminal\cite{Marchi2009} 
or Pfaffian wave functions\cite{Bajdich2006,Bajdich2008}.

\section{Acknowledgements}

We acknowledge financial support from the Cambridge Commonwealth Trust and 
UK Engineering and Physical Sciences Research Council (EPSRC)\@. 
The calculations were performed on the Cambridge High Performance 
Computing Service.

%

\end{document}